\documentclass[11pt]{article}

\usepackage[final]{acl}

\usepackage{times}
\usepackage{latexsym}

\usepackage[T1]{fontenc}

\usepackage[utf8]{inputenc}
\usepackage{amsfonts}
\usepackage{multirow}
\usepackage{array}
\usepackage{pifont}
\usepackage{algorithm}
\usepackage{algorithmic}
\usepackage{amssymb}
\usepackage{amsmath}
\usepackage{amsfonts}
\usepackage{footmisc}
\usepackage{multirow}
\usepackage{booktabs}
\usepackage[utf8]{inputenc}
\usepackage{url}
\usepackage{bbding}
\usepackage{wasysym}
\usepackage{utfsym}
\usepackage{xcolor,soul}
\usepackage{fontawesome}
\usepackage{hyperref}
\usepackage{array}
\usepackage{svg}
\usepackage{tikz}
\usepackage{circledsteps}
\usepackage{colortbl}
\usepackage[most]{tcolorbox}
\usepackage{lipsum}
\usepackage{enumitem}

\usepackage{microtype}
\usepackage{amsmath} 
\usepackage{multirow}
\usepackage[table,xcdraw]{xcolor}
\usepackage{inconsolata}
\usepackage{makecell}

\usepackage{graphicx}
\usepackage{url}
\usepackage{multirow}
\usepackage[table,xcdraw]{xcolor}
\usepackage{colortbl}
\usepackage[normalem]{ulem}
\usepackage{balance}
\useunder{\uline}{\ul}{}
\title{TellWhisper: Tell Whisper Who Speaks When}

\author{
  Yifan Hu$^{1,2}$, Peiji Yang$^2$, Zhisheng Wang$^2$, Yicheng Zhong$^2$, Rui Liu$^{1}$\thanks{Corresponding author.} \\
  $^1$~Inner Mongolia University, Hohhot, China \\
  $^2$~Tencent Technology Co.Ltd, Shenzhen, Guangdong, China \\
  \texttt{22309013@mail.imu.edu.cn, imucslr@imu.edu.cn},\\ 
  \texttt{\{peijiyang, plorywang, ajaxzhong\}@tencent.com}
}

\begin{document}
\maketitle
\begin{abstract}

Multi-speaker automatic speech recognition (MASR) aims to predict ``\textit{who spoke when and what}'' from multi-speaker speech, a key technology for multi-party dialogue understanding. However, most existing approaches decouple temporal modeling and speaker modeling when addressing ``\textit{when}'' and ``\textit{who}'': some inject speaker cues before encoding (e.g., speaker masking), which can cause irreversible information loss; others fuse identity by mixing speaker posteriors after encoding, which may entangle acoustic content with speaker identity. This separation is brittle under rapid turn-taking and overlapping speech, often leading to degraded performance.
To address these limitations, we propose \textbf{TellWhisper}, a unified framework that jointly models speaker identity and temporal within the speech encoder. Specifically, we design \textbf{TS-RoPE}, a time-speaker rotary positional encoding: time coordinates are derived from frame indices, while speaker coordinates are derived from speaker activity and pause cues. By applying region-specific rotation angles, the model explicitly captures per-speaker continuity, speaker-turn transitions, and state dynamics, enabling the attention mechanism to simultaneously attend to ``\textit{when}'' and ``\textit{who}''.
Moreover, to estimate frame-level speaker activity, we develop \textbf{Hyper-SD}, which casts speaker classification in hyperbolic space to enhance inter-class separation and refine speaker-activity estimates.
Extensive experiments demonstrate the effectiveness of the proposed approach. The project webpage is available at \url{https://walker-hyf.github.io/TellWhisper}.
\end{abstract}

\section{Introduction}
\begin{figure}[t]
\centering
\centerline{
\includegraphics[width=1\linewidth]{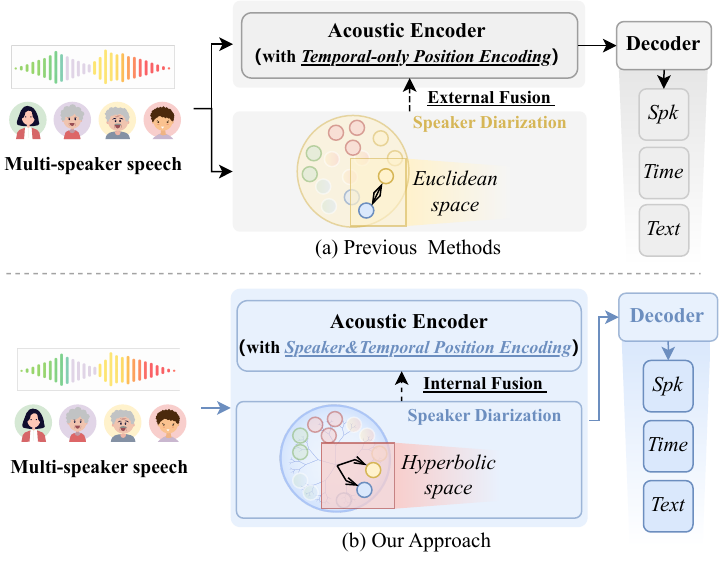}
}
\caption{(a) Prior methods model temporal structure and speaker information separately. (b) Our approach uses a unified positional encoding to capture both temporal and speaker dynamics.}
\label{fig:demo}
\end{figure}
Multi-speaker Automatic Speech Recognition (MASR) aims to predict who speaks what content and at what time in speech containing interactions among multiple speakers~\cite{polok2025dicow, yin2025speakerlm}.
It is a complex task that jointly integrates speaker diarization (SD)~\cite{bredin2020pyannote} and automatic speech recognition (ASR)~\cite{cao2012whisper}. 
With the development of speech intelligence and conversational systems, MASR plays an increasingly critical role in meeting and interview transcription~\cite{vinnikov2024notsofar}, multi-user human–computer interaction~\cite{shin2025enhancing}, and the construction of data for spoken dialogue speech foundation models~\cite{ju2025mooncast, xie2025soulx, hao2025mimo, weihua2026adamcot}.
Consequently, developing efficient and robust MASR models is of practical importance.

While current ASR models~\cite{yao2023zipformer, xu2025fireredasr} excel at recognizing linguistic content, their performance often degrades markedly in multi-party dialogues with rapid speaker-turn taking, largely because the critical cues of ``\textit{who}'' and ``\textit{when}'' remain insufficiently modeled.
In MASR, traditional solutions typically fuse SD and ASR outputs in parallel: the former predicts speaker identities and timestamps, the latter predicts content and timestamps, and the two streams are aligned by timestamps~\cite{yamasaki2023transcribing}. However, accurate timestamp alignment is challenging, especially under overlapping speech, and this pipeline often results in incorrect speaker assignment.
Recent works seek to unify SD and ASR, yet most approaches remain fundamentally factorized, modeling temporal structure and speaker identity separately and aggregating speaker cues with acoustic representations \emph{outside} the encoder.
As shown in Fig.~\ref{fig:demo}, they use absolute positional encoding for time modeling and adopt three common speaker strategies:
(1) ~\cite{polok2025dicow} masks non-target regions before encoding using SD labels, to preserve temporal, blank inputs are still decoded, which can trigger hallucinations.
(2) ~\cite{kang2025disentangling, he2025scaling} attempts to isolate the target speaker, but requires extra speaker prompts~\cite{ma2024extending, guo2024sq} or fixed number of separated individuals~\cite{zhao2023mossformer}, and struggles in overlapping regions.
(3) Other methods~\cite{park2024sortformer, medennikov2025streaming} add predefined speaker sinusoidal kernels weighted by posteriors to encoder states, such linear mixing entangles semantics with speaker cues and complicates decoding.
Therefore, how can we model temporal and speaker jointly \emph{within} the encoder in a more seamless way?

To overcome factorized modeling, we propose \textbf{TellWhisper} (Fig.~\ref{fig:demo}, lower). The model injects temporal and speaker information into the ASR encoder via positional encoding. Specifically, we design \textbf{TS-RoPE}, a time--speaker-aware rotary positional encoding, and apply it to encoder self-attention to modulate Query-Key dot products through controllable rotation-angle differences. 
We partition the Query/Key channels into temporal subspaces indexed by absolute frame time and speaker subspaces derived from per-frame activity to capture speaker-state dynamics (e.g., sustained speech and pauses). We also allocate disjoint channel regions to different speakers to avoid inter-speaker interference.
To obtain more reliable frame-level activity, we further propose \textbf{Hyper-SD}, which replaces Euclidean linear scoring with a hyperbolic ``feature-prototype distance''  (red box in Fig.~\ref{fig:demo}). Negative curvature induces exponential volume growth, so small shifts yield larger distance changes, improving separability among timbrally similar speakers and stabilizing speaker posteriors. 

In summary, the main contributions of this paper are as follows:
\begin{itemize}
    \item We propose TellWhisper, a novel multi-speaker ASR model that introduces TS-RoPE, a time--speaker-aware rotary positional encoding, into the speech encoder to naturally integrate temporal and speaker activity.
    \item To obtain reliable frame-level speaker activity, we develop Hyper-SD, a hyperbolic-space speaker diarization model that estimates speaker activity via ``feature-prototype distances.''
    \item We conduct extensive experiments that demonstrate the effectiveness of TS-RoPE for time-speaker integration and show that Hyper-SD provides reliable speaker-activity estimates.
\end{itemize}

\section{Related Works}
\subsection{Rotational Position Encoding}
\begin{figure*}[t]
\centering
\centerline{
\includegraphics[width=1\linewidth]{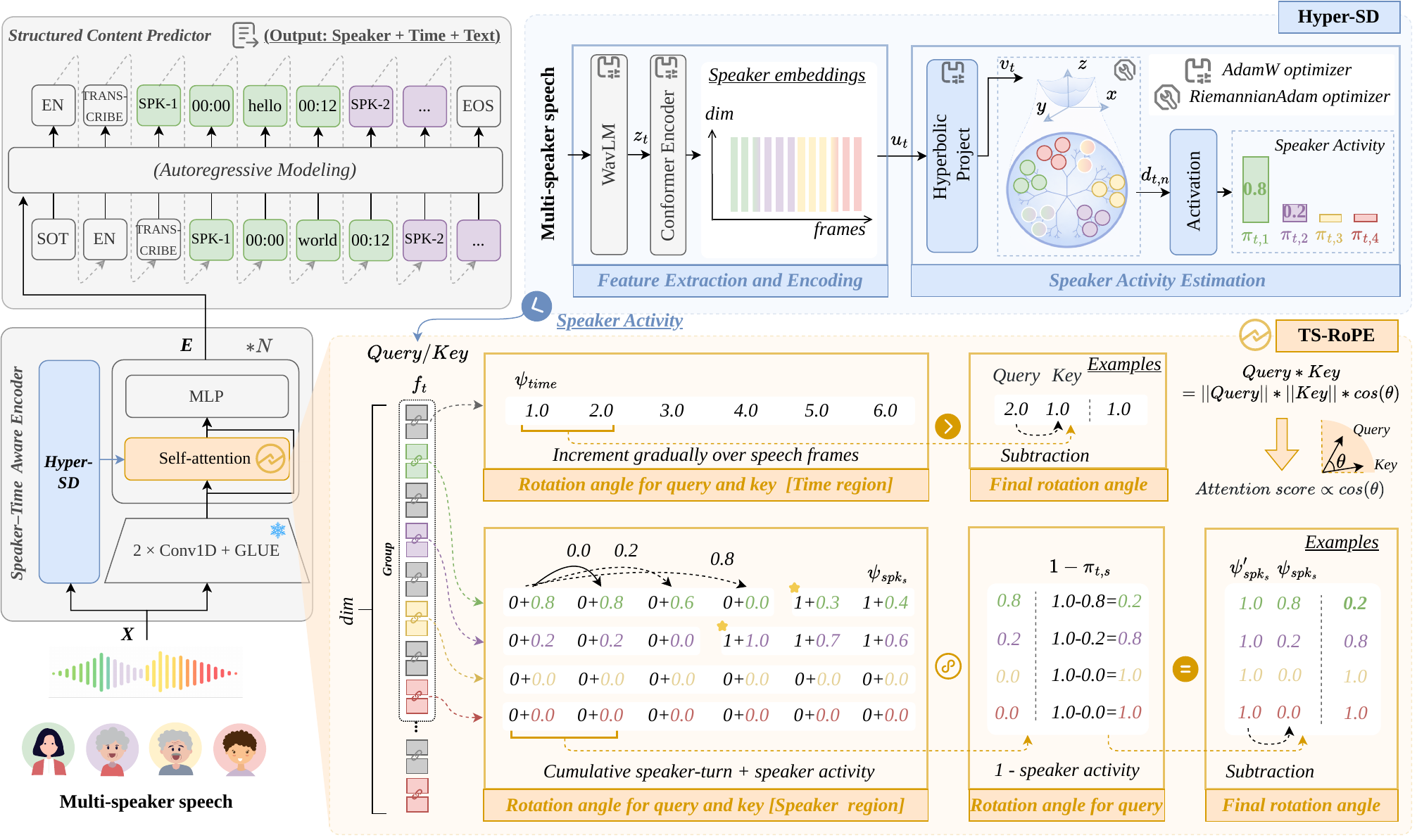}
}
\caption{
Overall architecture of the TellWhisper model. For multi-speaker speech, the Speaker-Time Aware Encoder encodes the input with convolutional layers and uses Hyper-SD to estimate frame-level speaker activity. Guided by TS-RoPE, self-attention jointly models temporal and speaker dynamics, and the Structured Content Predictor outputs speaker, time, and text. In particular, TS-RoPE builds separate temporal and speaker coordinates and encodes them into disjoint Query/Key subspaces, strengthening attention for aligning ``when'' and ``who'' cues.
}
\label{fig:asr_model}
\end{figure*}

Traditional absolute positional encoding (PE) injects fixed position-dependent vectors into semantic representations~\cite{vaswani2017attention}, requiring a predefined maximum length and failing to explicitly model relative positions. In contrast, Rotary Positional Embedding (RoPE)~\cite{su2024roformer} rotates Query and Key vectors so attention depends on relative angles, preserves norms, and supports long context.
Beyond large language models~\cite{bai2023qwen, touvron2023llama}, RoPE also applies to speech tasks such as ASR~\cite{zhang2025benchmarking} and speech enhancement~\cite{chen2024investigation}, where frame features rotate by time to encode dynamics. In vision, RoPE extends to multi-dimensional variants that encode multiple axes~\cite{lu2024fit}. More recently, multi-dimensional RoPE~\cite{yang2025qwen3} unifies positional encoding across modalities by partitioning channels into semantic subspaces (e.g., \textit{width} and \textit{height}) and encoding factors independently within shared attention.
Motivated by these advances, we target MASR, which requires joint temporal and speaker modeling. Instead of encoding time alone, we split channels into temporal and speaker subspaces: the temporal subspace uses standard time rotation, while the speaker subspace is modulated by speaker activity.


\subsection{Hyperbolic Representation Learning and Classification}
Conventional classifiers~\cite{bredin2020pyannote} typically use a linear head in Euclidean space, but Euclidean geometry’s flatness and polynomial volume growth make it hard to form large inter-class margins for highly similar distributions, leading to poor discrimination~\cite{xu2023hyperbolic}. Hyperbolic space, with negative curvature and exponential volume expansion, amplifies distance contrasts and enlarges margins~\cite{ganea2018hyperbolic}, which benefits speaker diarization where similar timbres produce confusable embeddings. Hyperbolic embeddings also capture hierarchical structure with low distortion~\cite{pal2024compositional} and use geometric cues (e.g., radius) to reflect a continuum from ambiguous to separable events~\cite{petermann2024hyperbolic}.
However, SD requires explicit frame-level discrete labels: non-speaking (noise/silence) should be grouped, while different overlap patterns (e.g., ``spk-A $\&$ spk-B'' vs.\ ``spk-B $\&$ spk-C'') must remain separable~\cite{bredin2020pyannote}. If ambiguous segments collapse near the origin, separability across overlap types degrades. Accordingly, we assign distinct labels to non-speaking segments, each single speaker, and each overlap combination, and enforce supervision that pushes features and prototypes toward well-separated boundary regions. Finally, we compute frame-level speaker activity from feature--prototype distances.

\section{Task Definition}
In multi-speaker automatic speech recognition, the input is a multi-speaker speech signal represented as a frame-level acoustic feature sequence $X=\{x_t\}_{t=1}^{T}$, where $x_t\in\mathbb{R}^{D}$ is the feature vector of frame $t$ and $T$ is the sequence length. The signal may contain overlap, rapid speaker transitions, and silence (non-speaking). The MASR model aims to infer structured outputs (speaker identities, timestamps, and transcribed text), formulated as
\begin{equation}
Y=\{(spk_s, [\tau_{s,j}^{\text{start}},\mathbf{y}_{s,j},\tau_{s,j}^{\text{end}})]^J_{j=1}\}_{s=1}^{S}
\end{equation}
where $spk_s$ denotes the speaker label, $\tau_{s,j}^{\text{start}}$ and $\tau_{s,j}^{\text{end}}$ are the segment boundaries for the $j$-th turn of speaker $s$, $\mathbf{y}_{s,j}$ is the associated text sequence, $J$ is the number of speaker-turn segments of $\mathrm{spk}_s$ in $X$, and $S$ is the number of speakers in $X$.

\section{Proposed Approach: TellWhisper}
As shown in Fig.~\ref{fig:asr_model}, we present the overall architecture of \textbf{TellWhisper}. We first describe how Hyper-SD estimates speaker activity, and then introduce the TS-RoPE-based time-speaker-aware encoder and the structured content predictor.

\subsection{Frame-level Speaker Activity Estimator}
As shown in the upper-right of Fig.~\ref{fig:asr_model}, Hyper-SD consists of two stages: (1) it learns speech representations from multiple WavLM~\cite{chen2022wavlm} layers and uses a Conformer encoder to inject global context into frame-level features; (2) a hyperbolic classifier maps Euclidean features into hyperbolic space and estimates speaker activity via feature--prototype distances.

\subsubsection{Speech Feature Extraction and Encoding}
Given multi-speaker speech $X$, we use WavLM to extract multi-layer frame-level representations $\mathbf{h}_t^{(l)}$.
A learnable weighted-sum aggregation fuses these features into a compact frame representation:
\begin{equation}
\mathbf{z}_t = \sum_{l=1}^{L} \alpha_l \mathbf{h}^{(l)}_t
\end{equation}
where, $l$ is the layer index, $t$ is the frame index, and $\alpha_l$ denotes the layer weight.

The aggregated features are then fed into a Conformer to model contextual dependencies:
\begin{equation}
\mathbf{u}_{1:T} = \mathrm{Conformer}(\mathbf{z}_{1:T})
\end{equation}
where $T$ is the number of frames. The Conformer integrates long-range context and local acoustic patterns to produce context-aware frame representations for speaker activity estimation.

\subsubsection{Prototype-Based Speaker Activity Estimation}
Speaker activity estimation is performed in hyperbolic space. Specifically, we first apply a linear transformation and norm clipping to the Euclidean feature $\mathbf{u}_t$:
\begin{equation}
\begin{gathered}
\mathbf{v}_t = \mathbf{W}\mathbf{u}_t + \mathbf{b} \in \mathbb{R}^{I},
\\
\mathbf{v}_t = \mathbf{v}_t \cdot
\min\!\left(1, \frac{r}{\lVert \mathbf{v}_t \rVert_2 + \epsilon}\right)
\end{gathered}
\end{equation}
Here, $I$ denotes the hyperbolic embedding dimension, $r$ controls the clipping radius, $\mathbf{W}$ and $\mathbf{b}$ are the weight matrix and bias, and $\epsilon$ is a small constant.

A Poincar\'e ball~\cite{ungar2001hyperbolic} $\mathbb{B}_{c}$ with curvature $c$ serves as the underlying hyperbolic space.
The clipped features are mapped to $\mathbb{B}_{c}$ via the exponential map at the origin and then projected to remain inside the ball for numerical stability.
We assign a learnable hyperbolic prototype $\mathbf{p}_n \in \mathbb{B}_{c}$ to each speaker-combination~\footnote{\emph{silence}; single-speaker sets $\{1\},\{2\},\{3\},\{4\}$; two-speaker overlaps $\{1,2\},\ldots,\{3,4\}$; three-speaker overlaps $\{1,2,3\},\ldots,\{2,3,4\}$; and $\{1,2,3,4\}$.} class $n \in \mathcal{N}$, where $\mathcal{N}$ is the power set of speakers and $|\mathcal{N}| = 2^{4}$ (we assume at most four speakers).
For each mapped frame-level embedding $\mathbf{v}^{'}_t$, we compute its hyperbolic distance to each prototype:
\begin{equation}
d_{t,n} = d_{\mathbb{B}_{c}}(\mathbf{v}^{'}_t, \mathbf{p}_n)
\end{equation}

Finally, the per-speaker frame-level activity $\pi_{t,s}$ is obtained by first applying an element-wise activation function to produce a joint distribution over all classes and then
marginalizing them:
\begin{equation}
\pi_{t,s}
= \sum_{n=1}^{2^\mathcal{N}} b_{s,n}\,\sigma(-d_{t,n}), s=1,2,3,4
\end{equation}
where $b_{s,n}\in\{0,1\}$ indicates whether speaker $s$ in class $n$.

\subsection{Speaker–Time Aware Encoder}
TellWhisper adopts TS-RoPE to inject temporal and frame-level speaker activity cues into self-attention by rotating Query/Key vectors in multiple interleaved rotary subspaces.

\subsubsection{Position Construction}
As shown in the lower-right part of Fig.~\ref{fig:asr_model}, for each encoder convolution layer output frame $f_t$, we construct a position vector consisting of one temporal index $\psi_{time}$ and four speaker-dependent indices $\psi_{spk_s}$. Meanwhile, we partition the $f_t$'s channel dimension $D$ into groups of 16 dimensions. Within each group, the 8 rotary pairs are assigned $\psi$ in an interleaved manner:
[$\psi_{time}$, $\psi_{spk_1}$, $\psi_{time}$, $\psi_{spk_2}$, $\psi_{time}$, $\psi_{spk_3}$, $\psi_{time}$, $\psi_{spk_4}$].
For the temporal position, we use the temporal index:
\begin{equation}
\psi_{time}(f_t)=t,\quad t\in\{0,1,\ldots,T-1\}
\end{equation}

For the speaker-dependent indices, to capture both \emph{within-speaker continuity} and \emph{speaker-turn}, we define a cumulative speaker-turn counts $\mathcal{C}$. It first obtain a binary activity indicator with a small threshold $\tau$ (e.g., if $\pi_{t-1,s}=0.03$ and $\pi_{t,s}=0.8$, then $a_{t-1,s}=0$ and $a_{t,s}=1$):
\begin{equation}
a_{t,s}=\mathbb{I}[\pi_{t,s}\geqslant \tau], \tau=0.1
\end{equation}

It then detect rising edges (i.e., a speaker starts speaking means a new turn segment / turn) and accumulate them:
\begin{equation}
\begin{gathered}
r_{t,s}=a_{t,s}(1-a_{t-1,s}),\quad a_{0,s}=0
\\
\mathcal{C}_{t,s}=\sum_{i=0}^{t} r_{i,s}
\end{gathered}
\end{equation}

Finally, the speaker position index is composed of the cumulative speaker-turn counts $C_{t,s}$ and a within-turn activity:
\begin{equation}
\psi_{spk_s}(f_t)=\mathcal{C}_{t,s}+\pi_{t,s}
\end{equation}

In addition, to encourage subsequent self-attention to focus more on the \emph{active-speaker} components in the Query, we introduce an extra, dynamic phase bias on the Query in speaker subspaces:
\begin{equation}
\begin{aligned}
\psi^{'}_{spk_s}(f_t) &= \psi_{{spk_s}}(f_t) + \bigl(1-\pi_{t,s}\bigr)
\end{aligned}
\end{equation}
note we apply the bias only to Query while keeping Key unchanged.

\subsubsection{TS-RoPE-Based Self-Attention}
Let $\mathbf{q}_{f_t'}, \mathbf{k}_{f_t} \in \mathbb{R}^{D}$ denote the Query and Key vectors at frame $f_t'$ and $f_t$. For the $i$-th rotary pair, if the pair in time region, the rotation angle is defined as:
\begin{equation}
    \theta_{f_t,i} = \psi_{time}(f_t)\,\omega_i, ~~\theta_{f_t',i} = \psi_{time}(f_t')\,\omega_i
\end{equation}
if the pair in speaker region:
\begin{equation}
    \theta_{f_t,i} =  \psi_{spk_s}(f_t)\,\omega_i, ~~\theta_{f_t',i} = \psi^{'}_{spk_{s}}(f_t')\,\omega_i
\end{equation}
where $\omega_i$ is the corresponding inverse frequency (all 8 rotary pairs within the same group share the same $\omega$):
\begin{equation}
\omega_i = \frac{1}{\mathrm{10000}^{\frac{2i}{D}}}, 
\quad i = 0, 1, \ldots, \frac{D}{16}-1
\end{equation}

The rotary transformation $\mathcal{R}$ is applied simultaneously to the Query and Key:
\begin{equation}
\begin{aligned}
\mathcal{R}(\mathbf{x}_{f_t})_i &=
\begin{bmatrix}
x_{f_t,2i} \cos \theta_{f_t,i} - x_{f_t,2i+1} \sin \theta_{f_t,i} \\
x_{f_t,2i} \sin \theta_{f_t,i} + x_{f_t,2i+1} \cos \theta_{f_t,i}
\end{bmatrix}, \\
&\quad \mathbf{x}_{f_t} \in \{\mathbf{q}_{f_t'}, \mathbf{k}_{f_t}\}
\end{aligned}
\end{equation}

After applying TS-RoPE, the attention weight between frames $f_t'$ and $f_{t}$ can be written as
\begin{equation}
    Attn(f_t',f_t) \propto
    \bigl\langle
    \mathcal{R}(\mathbf{q}_{f_t'}),
    \mathcal{R}(\mathbf{k}_{f_t})
    \bigr\rangle
\end{equation}

By coupling temporal positions with cumulative speaker phases, the resulting attention jointly captures temporal and speaker dynamics, yielding a fused representation $E$ that aligns ``\textit{who}'' and ``\textit{when}'' cues for the subsequent Structured Content Predictor.

\subsection{Structured Content Predictor}
As shown in the upper-left part of Fig.~\ref{fig:asr_model},
For the output content of TellWhisper, we adopt a segment-level structured modeling strategy. Specifically, temporally contiguous speech regions produced by the same speaker are treated as individual speech segments, each represented by an ordered sequence of tokens: $\langle spk_s \rangle$, $\langle t_{start} \rangle$, $\langle text \rangle$, and $\langle t_{end} \rangle$.
All speech segments from different speakers are concatenated in chronological order to form the final target sequence. For modeling, we employ a language-model-based autoregressive framework, treating the structured representation as a unified token sequence and training it using next-token prediction. During decoding, the model generates tokens sequentially conditioned on the encoded audio representations until the end-of-sequence token $\langle EOS \rangle$ is produced.

\section{Experiments}
To validate the effectiveness of the proposed TellWhisper in MASR task, we conduct comprehensive experiments. In addition, to assess the reliability of speaker activity produced by Hyper-SD, we carry out comprehensive evaluations on the SD task. In this section, we describe the experimental setup from the perspectives of Datasets, Metrics, Baseline Models and Training Strategy.

\subsection{Datasets}
\textbf{\textit{For the MASR task}}, we select four English multi-speaker datasets for training and evaluation. \textit{AMI} (SDM)~\cite{kraaij2005ami} and \textit{NotSoFar}~\cite{vinnikov2024notsofar} are collected from real-world multi-party meetings and recorded in far-field conditions, whereas \textit{Libri2Mix}~\cite{cosentino2020librimix} and \textit{LibriCSS}~\cite{chen2020continuous} are simulated. We also use single-utterance \textit{LibriSpeech}~\cite{panayotov2015librispeech} for preliminary fine-tuning before MASR training.
\textbf{\textit{For the SD task}}, we use six datasets for training and evaluation: \textit{AISHELL4}~\cite{fu2021aishell}, \textit{AliMeeting}~\cite{yu2022AliMeeting}, \textit{AMI}, \textit{MSDWild}~\cite{liu2022MSDWild}, \textit{RAMC}~\cite{yang2022ramc}, and \textit{VoxConverse}~\cite{chung2020VoxConverse}, all consisting of real-world multi-speaker conversations.
For detailed statistics (speech duration, overlap duration, and number of speakers), please refer to the Appendix~\ref{appendix:datasets}.

\subsection{Metrics}
To evaluate MASR in multi-speaker settings, conventional word error rate (\textit{WER}) is inadequate, as it fails to address speaker-permutation ambiguity and temporal misalignment. Using the Meeteval toolkit\footnote{https://github.com/fgnt/meeteval}, we report four metrics:
(1) \textbf{\textit{Concatenated minimum-permutation WER (CP-WER)}}, measuring content accuracy with speaker attribution.
(2) \textbf{\textit{Time-constrained minimum-permutation WER (TCP-WER)}}, adding temporal constraints to assess consistency of content, speaker, and time.
(3) \textbf{\textit{Optimal reference combination WER (ORC-WER)}}, a speaker-independent WER.
(4) \textbf{\textit{Time-constrained ORC-WER (TCORC-WER)}}, adding temporal constraints to \textit{ORC-WER}.
For \textit{TCP-WER} and \textit{TCORC-WER}, we set the collar to 0.5, i.e., a small forgiveness window around reference boundaries where timing deviations are ignored.

For SD, we use diarization error rate (\textit{DER}) with collar settings of 0.0 and 0.5.

\subsection{Baseline Models}
To comprehensively evaluate TellWhisper on the MASR task, we benchmark it against three categories of state-of-the-art baselines: (1) \textbf{\textit{Alignment-based models}}, including \textit{Pyannote3~\footnote{https://github.com/yinruiqing/pyannote-whisper}+Whisper} and \textit{Hyper-SD+Whisper}, which align and integrate the outputs of speaker diarization and a single-speaker ASR model via timestamps. (2) \textbf{\textit{Separation-based model}}, \textit{Tiger~\cite{xu2024tiger}+Whisper}, which first extracts the target speaker’s speech using the high-performing speech separation model and then performs single-speaker recognition. (3) \textbf{\textit{Single-stage prediction–based model}}, including \textit{Whisper-D} (fine-tuned directly from a single-utterance ASR model), \textit{SortFormer}~\cite{park2024sortformer} (adding speaker posteriors to the speech-encoder outputs), \textit{Dicow}~\cite{polok2025dicow} (applying speaker masks before speech encoding) and \textit{TellWhisper-Diarizen} (replace Hyper-SD with Diarizen). For a fair comparison, all baselines are trained and fine-tuned on the same backbone as \textit{TellWhisper}, i.e., Whisper large-v3-turbo~\footnote{\label{turbo}https://huggingface.co/openai/whisper-large-v3-turbo}.

To assess the reliability of Hyper-SD on the speaker diarization task, we compare it with two leading open-source models, \textit{Pyannote3}~\footnote{https://huggingface.co/pyannote/speaker-diarization-3.1} and \textit{Diarizen}~\cite{han2025leveraging}, both of which operate in Euclidean space. The former uses convolutional and linear layers, whereas the latter uses WavLM, Conformer, and a linear layer.

\subsection{Training Strategy}
We initialize TellWhisper with the pretrained Whisper large-v3-turbo~\footref{turbo} and freeze the first two convolutional layers of the encoder. To match Dicow’s training setup, we adopt a two-stage fine-tuning strategy: we first pre-fine-tune on single-speaker speech to learn structured content prediction for a single speaker, and then fine-tune on multi-speaker conversational speech to learn structured content prediction for multiple speakers. We apply the same training pipeline to Whisper-D and SortFormer. The models are trained with token-level cross-entropy using the AdamW optimizer~\cite{loshchilov2017adamw}.

For Hyper-SD, we initialize the WavLM backbone with WavLM-Large~\footnote{https://huggingface.co/microsoft/wavlm-large} and train on conversational data using NLLLoss. We optimize the hyperbolic classifier with RiemannianAdam~\cite{yun2023riemannian} and the remaining components with AdamW, employing a smaller learning rate for WavLM and a larger one for the other modules.

\section{Results and Discussions}

\begin{table}[h]
\centering
\resizebox{1\linewidth}{!}{
\begin{tabular}{c|cccccc}
\hline
 & \multicolumn{6}{c}{\textbf{DER ($\downarrow$)}} \\ \cline{2-7} 
 & $\zeta$=0s & \multicolumn{1}{c|}{$\zeta$=0.25s} & $\zeta$=0s & \multicolumn{1}{c|}{$\zeta$=0.25s} & $\zeta$=0s & $\zeta$=0.25s \\ \cline{2-7} 
\multirow{-3}{*}{Models} & \multicolumn{2}{c|}{AMI} & \multicolumn{2}{c|}{ AISHELL4} & \multicolumn{2}{c}{AliMeeting} \\ \hline
Pyannote3$^\blacktriangle$ & 22.60 & \multicolumn{1}{c|}{15.41} & 11.96 & \multicolumn{1}{c|}{6.27} & 24.40 & 15.67 \\
Diarizen$^\blacktriangle$ & {\ul 13.99} & \multicolumn{1}{c|}{{\ul 9.00}} & {\ul 9.94} & \multicolumn{1}{c|}{{\ul 4.78}} & {\ul 13.03} & {\ul 5.98} \\
\rowcolor[RGB]{250,239,239}
Hyper-SD & \textbf{13.62} & \multicolumn{1}{c|}{\textbf{8.82}} & \textbf{9.52} & \multicolumn{1}{c|}{\textbf{4.44}} & \textbf{10.76} & \textbf{4.59} \\ \hline
Models & \multicolumn{2}{c|}{MSDWild} & \multicolumn{2}{c|}{RAMC} & \multicolumn{2}{c}{VoxConverse} \\ \hline
Pyannote3$^\blacktriangle$ & 21.73 & \multicolumn{1}{c|}{12.25} & 20.91 & \multicolumn{1}{c|}{12.97} & 11.18 & 6.81 \\
Diarizen$^\blacktriangle$ & {\ul 12.33} & \multicolumn{1}{c|}{{\ul 5.09}} & {\ul 11.20} & \multicolumn{1}{c|}{{\ul 6.54}} & {\ul 9.19} & {\ul 5.74} \\
\rowcolor[RGB]{250,239,239}
Hyper-SD & \textbf{12.28} & \multicolumn{1}{c|}{\textbf{4.79}} & \textbf{10.94} & \multicolumn{1}{c|}{\textbf{6.48}} & \textbf{8.75} & \textbf{5.21} \\ \hline
\end{tabular}
}
\caption{\label{tab:result_sd}\textcolor{black}{Speaker diarization results of Hyper-SD on conversational speech. The symbol $\blacktriangle$ denotes models operating in Euclidean space. $\zeta$ is the collar.}} 
\end{table}

\begin{table*}[t]
\centering
\resizebox{0.9\linewidth}{!}{
\begin{tabular}{ccccccccc}
\hline
\multicolumn{1}{c|}{} & \multicolumn{4}{c|}{\textbf{CP-WER ($\downarrow$)}} & \multicolumn{4}{c}{\textbf{TCP-WER ($\downarrow$)}} \\ \cline{2-9} 
\multicolumn{1}{c|}{\multirow{-2}{*}{\textbf{Models}}} & Libri2Mix &AMI& NotSoFar  & \multicolumn{1}{c|}{LibriCSS} & Libri2Mix &AMI& NotSoFar  & LibriCSS \\ \hline
\rowcolor[RGB]{245,245,245} 
\multicolumn{9}{l}{\cellcolor[RGB]{245,245,245}{\textit{Processing: speaker diarization + single-speaker speech recognition (results alignment)}}} \\ \hline
\multicolumn{1}{c|}{Pyannote3+Whisper$^\P $} & 62.05 & 59.58 & 69.85 & \multicolumn{1}{c|}{44.34} & 62.08 & 61.21 & 70.89 & 44.74 \\
\multicolumn{1}{l|}{Hyper-SD+Whisper$^\P $} & {\color[HTML]{333333} 61.23} & {\color[HTML]{333333} 58.51} & {\color[HTML]{333333} 67.22} & \multicolumn{1}{c|}{{\color[HTML]{333333} 42.51}} & {\color[HTML]{333333} 61.25} & {\color[HTML]{333333} 59.62} & {\color[HTML]{333333} 67.84} & {\color[HTML]{333333} 42.68} \\ \hline
\rowcolor[RGB]{245,245,245} 
\multicolumn{9}{l}{\cellcolor[RGB]{245,245,245}\textit{Processing: speech decoupling $\rightarrow$ single-speaker speech recognition}} \\ \hline
\multicolumn{1}{c|}{Tiger+Whisper$^\P $} & 37.96 & - & - & \multicolumn{1}{c|}{-} & 37.97 & - & - & - \\ \hline
\rowcolor[RGB]{245,245,245} 
\multicolumn{9}{l}{\cellcolor[RGB]{245,245,245}\textit{Processing: multi-speaker speech recognition}} \\ \hline
\multicolumn{1}{c|}{Whisper-D$^\P $} & 14.48 & 35.23 & 38.04 & \multicolumn{1}{c|}{12.41} & 14.57 & 36.86 & 38.15 & 12.58 \\

\multicolumn{1}{c|}{SortFormer$^\P $} & {\color[HTML]{333333} 14.62} & {\color[HTML]{333333} 34.24} & {\color[HTML]{333333} 36.54} & \multicolumn{1}{c|}{{\color[HTML]{333333} 12.16}} & {\color[HTML]{333333} 14.76} & {\color[HTML]{333333} 35.96} & {\color[HTML]{333333} 36.73} & {\color[HTML]{333333} 12.88} \\

\multicolumn{1}{c|}{Dicow$^\P $} & \textbf{14.34} & {33.57} & {35.22} & \multicolumn{1}{c|}{{10.62}} & \textbf{14.35} & { 34.02} & {35.64} & {11.33} \\

\multicolumn{1}{c|}{TellWhisper-Diarizen} & {14.45} & {\ul 33.12} & {\ul 34.81} & \multicolumn{1}{c|}{{\ul 9.93}} & {14.87} & {\ul 33.72} & {\ul 34.86} & {\ul 11.15} \\

\rowcolor[RGB]{250,239,239} 
\multicolumn{1}{c|}{TellWhisper (ours)} & {\ul 14.39} & \textbf{32.53} & \textbf{34.48} & \multicolumn{1}{c|}{\textbf{9.88}} & {\ul 14.61} & \textbf{33.47} & \textbf{34.51} & \textbf{11.06} \\ \hline
\end{tabular}
}
\caption{\label{tab:result_masr_1}\textcolor{black}{Multi-speaker ASR results of TellWhisper on conversational speech. CP-WER measures content~+~speaker, TCP-WER measures time~+~content~+~speaker. The symbol $\P$ denotes absolute positional encoding.}}
\end{table*}

In this section, we comprehensively evaluate TellWhisper. We first validate the diarization capability of Hyper-SD and the reliability of its speaker-activity estimates. We then evaluate TellWhisper on MASR for jointly predicting speakers, timestamps, and transcribed content. To quantify the contribution of each TS-RoPE component, we conduct ablation studies. Finally, we visualize the distribution of Hyper-SD class prototypes in hyperbolic space. In Appendix~\ref{appendix:case}, we further provide qualitative case studies on the impact of Hyper-SD's curvature hyperparameter $c$ on classification performance, as well as TellWhisper's recognition performance under different overlap ratios.

\begin{table}[]
\centering
\resizebox{1\linewidth}{!}{
\begin{tabular}{c|cccc}
\hline
 & \multicolumn{4}{c}{OCR-WER ($\downarrow$)} \\ \cline{2-5} 
\multirow{-2}{*}{Models} & Libri2Mix &AMI& NotSoFar  & LibriCSS \\ \hline
Whisper-D$^\P $ & 14.39 & 34.16 & 35.67 & 11.96 \\
SortFormer$^\P $ & {\color[HTML]{333333} 14.51} & {\color[HTML]{333333} 33.11} & {\color[HTML]{333333} 34.52} & {\color[HTML]{333333} 11.73} \\
Dicow$^\P $ & {\ul 13.34} & {32.83} & \textbf{32.20} & {9.43} \\
TellWhisper-Diarizen & {13.46} & {\ul 31.35} & {32.52} & {\ul 9.16} \\
\rowcolor[RGB]{250,239,239} 
\textbf{TellWhisper (ours)} & \textbf{13.32} & \textbf{30.72} & {\ul 32.31} & \textbf{9.14} \\ \hline
 & \multicolumn{4}{c}{TCORC-WER ($\downarrow$)} \\ \cline{2-5} 
\multirow{-2}{*}{Models} & Libri2Mix &AMI& NotSoFar  & LibriCSS \\ \hline
Whisper-D$^\P $ & 14.40 & 35.81 & 34.24 & 12.25 \\
SortFormer$^\P $ & {\color[HTML]{333333} 14.55} & {\color[HTML]{333333} 34.57} & {\color[HTML]{333333} 35.21} & {\color[HTML]{333333} 12.42} \\
Dicow$^\P $ & \textbf{13.36} & {33.53} & {\ul 32.43} & { 11.05} \\
TellWhisper-Diarizen & {13.83} & {\ul 32.11} & {32.45} & {\ul 10.47} \\
\rowcolor[RGB]{250,239,239}
\textbf{TellWhisper (ours)} & {\ul 13.67} & \textbf{31.87} & \textbf{32.36} & \textbf{10.42} \\ \hline
\end{tabular}
}
\caption{\label{tab:result_masr_2}\textcolor{black}{Multi-speaker ASR results of TellWhisper on conversational speech. CP-WER measures content, TCP-WER measures time~+~content. The symbol $\P$ denotes absolute positional encoding.}}
\end{table}

\subsection{Verifying the Reliability of Hyper-SD}
In this experiment, we compare against Pyannote3 and Diarizen. Table~\ref{tab:result_sd} reports DER under two collar settings (0 s and 0.25 s). Overall, Hyper-SD attains the best DER on all datasets for both collars, indicating robust and consistent gains. In particular, both Diarizen and Hyper-SD markedly outperform Pyannote3, indicating that WavLM-based encoders can extract richer speaker-related acoustic information from speech frames than CNN-based structure. Compared with Diarizen, Hyper-SD yields the largest improvement on AliMeeting (the improvement is 2.27 when c~=~0 s and 1.59 when c~=~0.25 s), indicating more robust speaker separability and activity estimation in challenging real meeting conditions. Consistent improvements are also observed on other datasets, e.g.,AMI(13.99 $\rightarrow$ 13.62; 9.00 $\rightarrow$ 8.82) and AISHELL4 (9.94 $\rightarrow$ 9.52; 4.78 $\rightarrow$ 4.44). These results indicate that classifying learned speech representations in hyperbolic space is more effective than performing linear classification directly in Euclidean space. This further supports the reliability of its speaker-activity estimation, providing a more stable prior for subsequent ``who speaks when'' modeling in MASR.

\begin{table*}[t]
\centering
\resizebox{0.9\linewidth}{!}{
\begin{tabular}{lcccccccc}
\hline
\multicolumn{1}{c|}{} & \multicolumn{4}{c|}{\textbf{CP-WER ($\downarrow$)}} & \multicolumn{4}{c}{\textbf{TCP-WER ($\downarrow$)}} \\ \cline{2-9} 
\multicolumn{1}{c|}{\multirow{-2}{*}{Models}} & Libri2Mix &AMI& NotSoFar  & \multicolumn{1}{c|}{LibriCSS} & Libri2Mix &AMI& NotSoFar  & LibriCSS \\ \hline
\rowcolor[RGB]{250,239,239}
\multicolumn{1}{l|}{TellWhisper ($\textcircled{A})$} & \textbf{14.39} & \textbf{32.53} & \textbf{34.48} & \multicolumn{1}{c|}{\textbf{9.88}} &\textbf{14.61} &\textbf{33.47} & \textbf{34.51} & \textbf{11.06} \\ \hline
\multicolumn{1}{l|}{$\textcircled{A}$~w/o M$\_query$ ($\textcircled{B}$)} & {\ul 15.13} & {\ul 35.02} & {\ul 36.27} & \multicolumn{1}{c|}{{\ul 10.82}} & {\ul 15.38} & {\ul 35.26} & {\ul 37.13} & {\ul 12.61} \\ \hline
\multicolumn{1}{l|}{$\textcircled{B}$~w/o M$\_{speaker\text{-}turn}$ ($$\textcircled{C}$$)} & 15.53 & 36.22 & 38.13 & \multicolumn{1}{c|}{11.68} & 15.60 & 36.68 & 39.23 & 12.84 \\ \hline
\multicolumn{1}{l|}{$\textcircled{C}$~w/o M$\_activity$} & 15.48 & 36.84 & 39.54 & \multicolumn{1}{c|}{12.32} & 15.50 & 36.89 & 39.63 & 12.75 \\ \hline
\end{tabular}
}
\caption{\label{tab:result_masr_3}\textcolor{black}{Ablation results of TellWhisper, where $M_{\text{query}}$ denotes the extra angular rotation applied to the Query speaker region, $M_{\text{speaker-turn}}$ denotes cumulative speaker-turn counts, and $M_{\text{activity}}$ denotes speaker activity.
}}
\end{table*}

\subsection{Evaluating the Performance of Multi-Speaker Speech Recognition}
In the MASR experiments, we evaluate on four datasets, and the results in Table~\ref{tab:result_masr_1} exhibit a clear hierarchy across paradigms. The ``diarization + single-speaker ASR'' pipeline performs worst, indicating strong sensitivity to upstream separation/alignment errors and error propagation. Tiger+Whisper reduces Libri2Mix WER to 37.96/37.97, yet still falls behind direct multi-speaker recognition. Among single-stage systems, TellWhisper achieves the best performance and TellWhisper-Diarizen the second-best on AMI, NotSoFar, and LibriCSS, consistently surpassing Dicow while also reducing TCP-WER, suggesting improved speaker attribution without compromising timestamp accuracy. TellWhisper further outperforms TellWhisper-Diarizen on all datasets (e.g., WER $-0.59/-0.25$ on AMI), confirming the benefit of Hyper-SD. On fully overlapped Libri2Mix, our approach matches the strongest baseline, with larger gains on real meetings. This is likely due to Libri2Mix’s construction: overlap starts at time zero and each speaker has a single utterance, resulting in no speaker-turn transitions. As TS-RoPE targets speaker-aware temporal dynamics, such structure offers limited headroom for further WER reductions, while remaining competitive under extreme overlap.

Table~\ref{tab:result_masr_2} further corroborates this conclusion from a content-centric perspective: TellWhisper reduces OCR-WER to 30.72~/~9.14 on AMI~/~LibriCSS and achieves the lowest TCOCR-WER on AMI~/~NotSoFar ~/~LibriCSS (31.87~/~32.36~/~10.42), with only a slight degradation relative to Dicow on Libri2Mix. Overall, TellWhisper’s advantages are most evident in real meeting and conversational scenarios with more frequent overlap and more complex speaker turns, demonstrating stronger speaker modeling and more robust temporal alignment.

\subsection{Ablation Results}
We ablate the design of speaker-region positional indices in TS-RoPE. As shown in Table~\ref{tab:result_masr_3}, with all components enabled, TellWhisper achieves optimal performance on both CP-WER and TCP-WER. 
Removing the extra Query-side phase bias (w/o $M_{\text{query}}$) consistently degrades performance (CP-WER +0.74$\backsim$2.49; TCP-WER +0.77$\backsim$2.62), suggesting this Query-only phase encourages attention to emphasize active speakers, improving speaker assignment and temporal alignment.
Further removing the cumulative speaker-turn counts (w/o $M_{\text{speaker\mbox{-}turn}}$) causes larger drops (CP-WER +1.14$\backsim$3.69; TCP-WER +0.99$\backsim$4.72), especially on AMI/NotSoFar , highlighting the importance of cumulative turn information for continuity and turn boundaries.
When removing posterior-based activity cues in the speaker region (w/o $M_{\text{posterior}}$), performance drops most severely (NotSoFar  CP-WER~/~TCP-WER +5.06~/~+5.12), indicating posteriors are the key signal for identifying active speakers and maintaining stable alignment.

\subsection{Visualization Results}
\begin{figure}[t]
\centering
\centerline{
\includegraphics[width=1\linewidth]{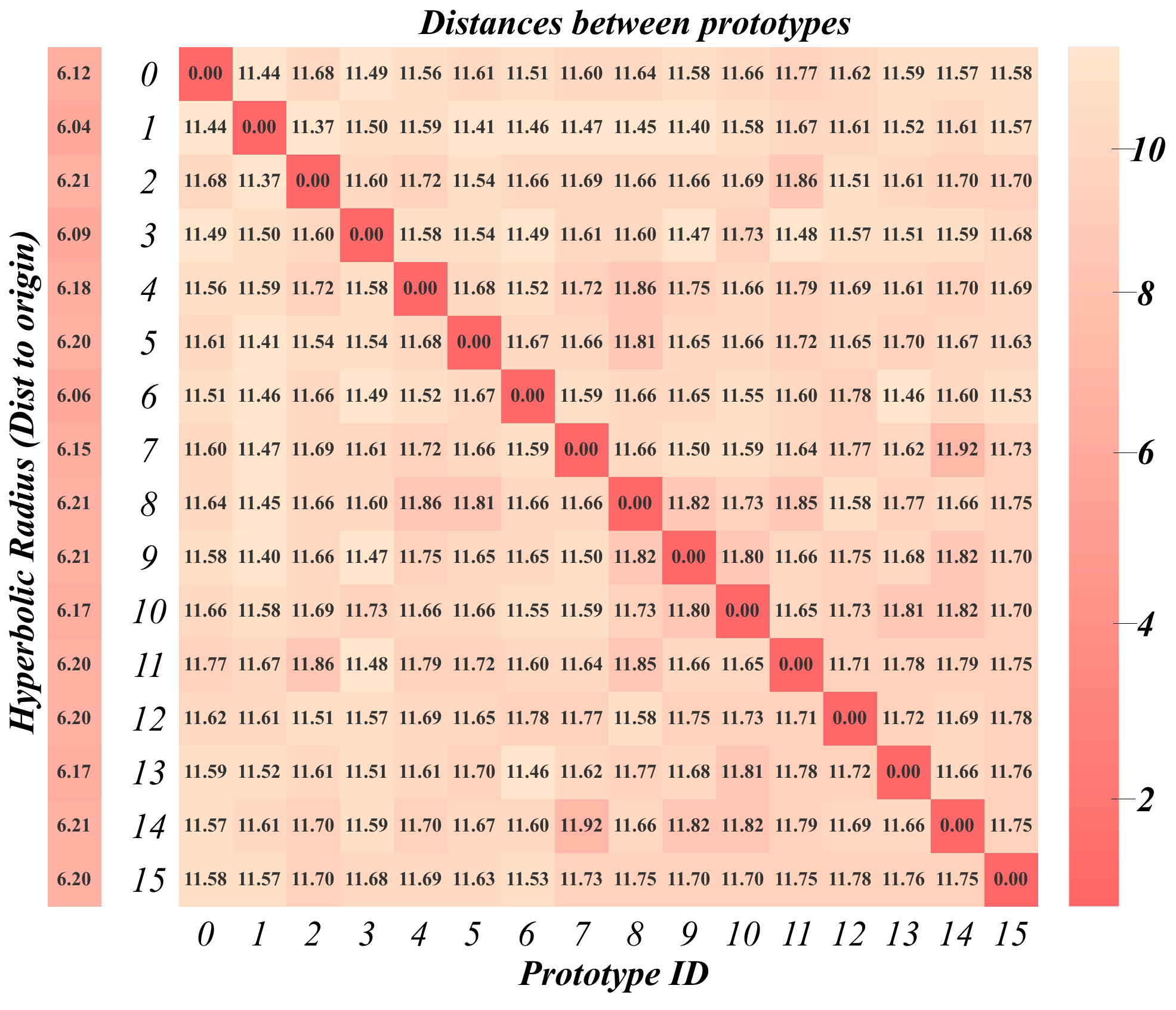}
}
\caption{Visualization of hyperbolic distances among the 16 class prototypes and their distances to the origin in hyperbolic-space-based speaker activity estimation.}
\label{fig:prototypes}
\end{figure}

As SD requires frame-level assignment to speaker classes, it primarily relies on fine-grained discriminative structure rather than an abstract-to-specific hierarchy. We therefore visualize the learned prototypes by plotting their pairwise hyperbolic distance matrix together with each prototype’s radial distance to the origin. As shown in Fig.~\ref{fig:prototypes}, the inter-prototype distances are largely uniform (around 11-12, right) and the radii vary within a narrow range (around 6.0-6.2, left), 
indicating that the prototypes are well separated and exhibit no clear hierarchical stratification.

\section{Conclusion}
We present TellWhisper, a unified framework for multi-speaker automatic speech recognition that couples temporal structure with speaker dynamics in the speech encoder. The core of TellWhisper is TS-RoPE, a time-speaker-aware rotary encoding that partitions Query/Key channels into temporal and speaker subspaces and applies region-specific rotations to align ``when'' and ``who'' cues in self-attention.
TS-RoPE uses frame-level speaker activity to build speaker coordinates that capture within-speaker continuity and turn transitions. For reliable activity estimates, Hyper-SD performs prototype-based speaker-combination classification in hyperbolic space and derives activity from feature-prototype distances.
Experiments show TellWhisper improves recognition accuracy, speaker attribution, and time consistency, while Hyper-SD delivers robust diarization and stable activity priors. These results indicate time-speaker-aware positional modeling and geometry-aware classification effectively support multi-speaker speech understanding.


\section*{Limitations}
\textbf{(1) The number of supported speakers is limited.} In this work, TS-RoPE is designed for 1-4 speaker scenarios, consistent with the settings in SortFormer and DiCoW. We further extend the model to accommodate more speakers in future work. \textbf{(2) The geometric consistency of Hyper-SD remains to be improved.} At present, we perform hyperbolic classification only after feature extraction, leaving the encoder and the hyperbolic classifier in mismatched embedding spaces. In future work, we investigate end-to-end hyperbolic learning that encodes features directly in hyperbolic space using hyperbolic neural networks.

\section*{Ethics Statement}
\textbf{(1) About the datasets}: All datasets used in this work are sourced from publicly available corpora and are accessed only after obtaining approval through the official application processes on their respective websites. We use only the audio recordings and the transcripts, and the data do not contain personally identifying information or offensive content.
\textbf{(2) About AI assistants}: During model development, we use AI tools to assist with code writing, and all AI-generated code is subsequently reviewed and verified by humans. During paper writing, we use AI only for grammar checking.

\section*{Acknowledgment}
This research of Rui Liu was funded by the General Program (No.62476146) of the National Natural Science Foundation of China, the Young Elite Scientists Sponsorship Program by CAST (2024QNRC001), the Outstanding Youth Project of Inner Mongolia Natural Science Foundation (2025JQ011), the Key R\&D and Achievement Transformation Program of Inner Mongolia Autonomous Region (2025YFHH0014), the Central Government Fund for Promoting Local Scientific and Technological Development (2025ZY0143).

\bibliography{custom}


\appendix
\section*{Technical Appendix}
\label{sec:appendix}
In this technical appendix, we provide additional details of TellWhisper for reference, including experimental settings and supplementary results.

\section{More Details of Experiments}
\label{sec:appendix_experiments}
In this section, we provide additional experimental details, including the datasets, experimental setup.

\begin{table}[h]
\centering
\resizebox{1\linewidth}{!}{
\begin{tabular}{c|c|c|c|c}
\hline
\textbf{Datasets} &
\textbf{Split} &
\makecell{\textbf{Speech}\\\textbf{Duration}} &
\makecell{\textbf{Overlap}\\\textbf{Duration}} &
\makecell{\textbf{Max}\\\textbf{Speaker}}
 \\ \hline
\multirow{3}{*}{AMI} & train & 65.81 & 8.59 & 4 \\
 & dev & 7.69 & 1.06 & 4 \\
 & test & 7.39 & 1.04 & 4 \\ \hline
\multirow{3}{*}{NotSoFar } & train & 31.15 & 6.80 & 4 \\
 & dev & 13.99 & 3.51 & 4 \\
 & test & 15.99 & 3.95 & 4 \\ \hline
\multirow{3}{*}{Libri2Mix} & train & 346.88 & 264.82 & 2 \\
 & dev & 7.23 & 4.21 & 2 \\
 & test & 2.16 & 1.42 & 2 \\ \hline
\multirow{2}{*}{LibriCSS} & dev & 1.00 & 0.07 & 4 \\
 & test & 8.66 & 0.60 & 4 \\ \hline
\end{tabular}
}
\caption{\label{tab:statistic_data}\textcolor{black}{Statistics of the MASR datasets, including speech duration (\textit{h}), overlapped-speech duration (\textit{h}), and the maximum number of speakers.}} 
\end{table}

\begin{table}[h]
\centering
\resizebox{1\linewidth}{!}{
\begin{tabular}{c|c|cccc}
\hline
\multirow{2}{*}{\textbf{Datasets}} & \multirow{2}{*}{\textbf{Split}} & \multicolumn{4}{c}{\textbf{Speaker proportion}} \\ \cline{3-6} 
 &  & \textbf{1} & \textbf{2} & \textbf{3} & \textbf{4} \\ \hline
\multirow{3}{*}{AMI} & train & 12.67 & 24.75 & 33.51 & 29.07 \\
 & dev & 12.41 & 21.75 & 30.85 & 34.99 \\
 & test & 14.59 & 23.09 & 32.36 & 29.96 \\ \hline
\multirow{3}{*}{NotSoFar } & train & 1.95 & 6.56 & 17.97 & 73.52 \\
 & dev & 2.19 & 8.11 & 16.43 & 73.25 \\
 & test & 3.83 & 8.35 & 24.51 & 63.31 \\ \hline
\multirow{3}{*}{Libri2Mix} & train & 0.00 & 100.00 & 0.00 & 0.00 \\
 & dev & 0.00 & 100.00 & 0.00 & 0.00 \\
 & test & 0.00 & 100.00 & 0.00 & 0.00 \\ \hline
\multirow{2}{*}{LibriCSS} & dev & 10.64 & 29.13 & 30.48 & 29.75 \\
 & test & 11.85 & 28.64 & 30.68 & 28.83 \\ \hline
\end{tabular}
}
\caption{\label{tab:statistic_spk}\textcolor{black}{Speaker-count distribution of the multi-speaker ASR datasets, reporting the proportion (\textit{\%}) of utterances with each number of speakers in each dataset.}} 
\end{table}

\begin{table}[h]
\centering
\resizebox{1\linewidth}{!}{
\begin{tabular}{c|c|c|c|c}
\hline
{\color[HTML]{333333} \textbf{Datasets}}             & {\color[HTML]{333333} \textbf{Split}} & {\color[HTML]{333333} \textbf{\begin{tabular}[c]{@{}c@{}}Speech\\ duration\end{tabular}}} & {\color[HTML]{333333} \textbf{\begin{tabular}[c]{@{}c@{}}Overlap\\ duration\end{tabular}}} & {\color[HTML]{333333} \textbf{\begin{tabular}[c]{@{}c@{}}Max\\speaker\end{tabular}}} \\ \hline
{\color[HTML]{333333} }                              & {\color[HTML]{333333} train}          & {\color[HTML]{333333} 97.22}                                                              & {\color[HTML]{333333} 87.44}                                                               & {\color[HTML]{333333} 7}                    \\
{\color[HTML]{333333} }                              & {\color[HTML]{333333} dev}            & {\color[HTML]{333333} 9.36}                                                               & {\color[HTML]{333333} 0.76}                                                                & {\color[HTML]{333333} 7}                    \\
\multirow{-3}{*}{{\color[HTML]{333333} AISHELL4}}    & {\color[HTML]{333333} test}           & {\color[HTML]{333333} 11.51}                                                              & {\color[HTML]{333333} 0.57}                                                                & {\color[HTML]{333333} 7}                    \\ \hline
{\color[HTML]{333333} }                              & {\color[HTML]{333333} train}          & {\color[HTML]{333333} 64.98}                                                              & {\color[HTML]{333333} 8.72}                                                                & {\color[HTML]{333333} 5}                    \\
{\color[HTML]{333333} }                              & {\color[HTML]{333333} dev}            & {\color[HTML]{333333} 7.00}                                                               & {\color[HTML]{333333} 0.99}                                                                & {\color[HTML]{333333} 4}                    \\
\multirow{-3}{*}{{\color[HTML]{333333} AMI}}         & {\color[HTML]{333333} test}           & {\color[HTML]{333333} 7.29}                                                               & {\color[HTML]{333333} 1.06}                                                                & {\color[HTML]{333333} 4}                    \\ \hline
{\color[HTML]{333333} }                              & {\color[HTML]{333333} train}          & {\color[HTML]{333333} 103.44}                                                             & {\color[HTML]{333333} 29.71}                                                               & {\color[HTML]{333333} 4}                    \\
{\color[HTML]{333333} }                              & {\color[HTML]{333333} dev}            & {\color[HTML]{333333} 3.88}                                                               & {\color[HTML]{333333} 0.84}                                                                & {\color[HTML]{333333} 4}                    \\
\multirow{-3}{*}{{\color[HTML]{333333} AliMeeting}}  & {\color[HTML]{333333} test}           & {\color[HTML]{333333} 9.91}                                                               & {\color[HTML]{333333} 2.02}                                                                & {\color[HTML]{333333} 4}                    \\ \hline
{\color[HTML]{333333} }                              & {\color[HTML]{333333} train}          & {\color[HTML]{333333} 58.67}                                                              & {\color[HTML]{333333} 6.84}                                                                & {\color[HTML]{333333} 10}                   \\
{\color[HTML]{333333} }                              & {\color[HTML]{333333} dev}            & {\color[HTML]{333333} 6.15}                                                               & {\color[HTML]{333333} 0.72}                                                                & {\color[HTML]{333333} 7}                    \\
\multirow{-3}{*}{{\color[HTML]{333333} MSDWild}}     & {\color[HTML]{333333} test}           & {\color[HTML]{333333} 7.07}                                                               & {\color[HTML]{333333} 0.76}                                                                & {\color[HTML]{333333} 9}                    \\ \hline
{\color[HTML]{333333} }                              & {\color[HTML]{333333} train}          & {\color[HTML]{333333} 128.68}                                                             & {\color[HTML]{333333} 1.20}                                                                & {\color[HTML]{333333} 10}                   \\
{\color[HTML]{333333} }                              & {\color[HTML]{333333} dev}            & {\color[HTML]{333333} 8.23}                                                               & {\color[HTML]{333333} 0.04}                                                                & {\color[HTML]{333333} 2}                    \\
\multirow{-3}{*}{{\color[HTML]{333333} RAMC}}        & {\color[HTML]{333333} test}           & {\color[HTML]{333333} 17.19}                                                              & {\color[HTML]{333333} 0.14}                                                                & {\color[HTML]{333333} 2}                    \\ \hline
{\color[HTML]{333333} }                              & {\color[HTML]{333333} train}          & {\color[HTML]{333333} 16.98}                                                              & {\color[HTML]{333333} 0.63}                                                                & {\color[HTML]{333333} 20}                   \\
{\color[HTML]{333333} }                              & {\color[HTML]{333333} dev}            & {\color[HTML]{333333} 1.93}                                                               & {\color[HTML]{333333} 0.08}                                                                & {\color[HTML]{333333} 15}                   \\
\multirow{-3}{*}{{\color[HTML]{333333} VoxConverse}} & {\color[HTML]{333333} test}           & {\color[HTML]{333333} 38.99}                                                              & {\color[HTML]{333333} 1.19}                                                                & {\color[HTML]{333333} 21}                   \\ \hline
\end{tabular}
}
\caption{\label{tab:statistic_sd}\textcolor{black}{Statistics of the speaker diarization datasets, including speech duration, overlapped-speech duration, and the maximum number of speakers.}} 
\end{table}

\subsection{Datasets}
\label{appendix:datasets}
Statistics of the four MASR datasets are summarized in Table~\ref{tab:statistic_data}, including the duration breakdown of the training, validation, and test splits, as well as the proportion of overlapping speech in each dataset. Among them, Libri2Mix exhibits the highest overlap ratio, mainly because each utterance is constructed by mixing two single-sentence recordings from different speakers, resulting in overlap starting from time \textit{0:00}. In addition, to match the TS-RoPE setting in our model, we segment all datasets such that each utterance contains at most four speakers (i.e., 1-4 speakers). As shown in Table~\ref{tab:statistic_spk}, each dataset includes multi-speaker utterances with different speaker-count distributions.

Statistics of the six SD datasets are reported in Table~\ref{tab:statistic_sd}, including total speech duration, overlapping speech duration, and the maximum number of speakers. During Hyper-SD training, we store time-stamped supervision in \textit{RTTM} format. Each training chunk contains 799 frames, and we additionally impose an upper bound on the number of speakers per segment (i.e., 4 speakers).

\subsection{Experimental Setup}
As shown in Table~\ref{tab:statistic_model}, we report the key hyperparameters of the main modules in TellWhisper. During training, we adopt different optimizers and learning rates for different components. 
(1) \textbf{\textit{Speaker Activity Estimation (Hyper-SD).}} Optimizing parameters in hyperbolic space is a manifold-constrained problem with curvature, where standard Adam/AdamW (which performs Euclidean gradient updates) may lead to incorrect update directions, drifting off the manifold, and numerical instability. Therefore, for the hyperbolic speaker prototypes, we use Riemannian Adam, which performs Adam-style updates on the hyperbolic manifold, resulting in more stable optimization and faster convergence. The learning rate is set to $1\times10^{-3}$. For the WavLM parameters, we use AdamW with a learning rate of $2\times10^{-5}$; all remaining parameters are optimized with AdamW using a learning rate of $1\times10^{-3}$. 
(2) \textbf{\textit{Speaker-Time Aware Encoder}} and \textbf{\textit{Structured Content Predictor.}} We use AdamW with a learning rate of $1\times10^{-5}$ and $\epsilon=1\times10^{-8}$. 
All experiments are conducted on eight NVIDIA H20 GPUs.

\begin{table}[t]
\centering
\resizebox{1\linewidth}{!}{
\begin{tabular}{ccc}
\hline
\multicolumn{1}{c|}{\textbf{Module}} & \multicolumn{1}{c|}{\textbf{Hyperparameter}} & \textbf{Value} \\ \hline
\multicolumn{3}{l}{\cellcolor[HTML]{EFEFEF}\textit{Frame-level Speaker Activity Estimator (Hyper-SD)}} \\ \hline
\multicolumn{1}{c|}{} & \multicolumn{1}{c|}{wavlm\_layer\_num} & 25 \\
\multicolumn{1}{c|}{\multirow{-2}{*}{WavLM}} & \multicolumn{1}{c|}{wavlm\_feat\_dim} & 1024 \\ \hline
\multicolumn{1}{c|}{} & \multicolumn{1}{c|}{attention\_in} & 256 \\
\multicolumn{1}{c|}{} & \multicolumn{1}{c|}{num\_head} & 4 \\
\multicolumn{1}{c|}{\multirow{-3}{*}{Conformer}} & \multicolumn{1}{c|}{use\_posi} & false \\ \hline
\multicolumn{1}{c|}{} & \multicolumn{1}{c|}{input\_dim} & 256 \\
\multicolumn{1}{c|}{\multirow{-2}{*}{\begin{tabular}[c]{@{}c@{}}Hyperbolic\\ Projection\end{tabular}}} & \multicolumn{1}{c|}{output\_dim} & 128 \\ \hline
\multicolumn{1}{c|}{} & \multicolumn{1}{c|}{hyperbolic\_dim} & 128 \\
\multicolumn{1}{c|}{} & \multicolumn{1}{c|}{margin} & 0.3 \\
\multicolumn{1}{c|}{\multirow{-3}{*}{\begin{tabular}[c]{@{}c@{}}Hyperbolic\\ classifier\end{tabular}}} & \multicolumn{1}{c|}{num\_classes} & 16 \\ \hline

\multicolumn{3}{l}{\cellcolor[HTML]{EFEFEF}\textit{Speaker–Time Aware Encoder}} \\ \hline
\multicolumn{1}{c|}{} & 
\multicolumn{1}{c|}{text\_n\_vocab} & 51866 \\
\multicolumn{1}{c|}{} & 
\multicolumn{1}{c|}{speech\_sample\_rate} & 16000 \\
\multicolumn{1}{c|}{\multirow{-3}{*}{Tokenizer}} & 
\multicolumn{1}{c|}{speech\_n\_mels} & 128 \\ \hline

\multicolumn{1}{c|}{} & \multicolumn{1}{c|}{d\_model} & 1280 \\
\multicolumn{1}{c|}{} & \multicolumn{1}{c|}{attention\_heads} & 20 \\
\multicolumn{1}{c|}{} & \multicolumn{1}{c|}{speaker\_activity} & 0-1 \\
\multicolumn{1}{c|}{} & \multicolumn{1}{c|}{T} & 1500 \\
\multicolumn{1}{c|}{} & \multicolumn{1}{c|}{ffn\_dim} & 5120 \\

\multicolumn{1}{c|}{\multirow{-6}{*}{Self-Attention+MLP}} & \multicolumn{1}{c|}{layers (N)} & 32 \\ \hline
\multicolumn{3}{l}{\cellcolor[HTML]{EFEFEF}\textit{Structured Content Predictor}} \\ \hline
\multicolumn{1}{c|}{} & \multicolumn{1}{c|}{attention\_heads} & 20 \\
\multicolumn{1}{c|}{} & \multicolumn{1}{c|}{ffn\_dim} & 5120 \\
\multicolumn{1}{c|}{} & \multicolumn{1}{c|}{layers} & 4 \\
\multicolumn{1}{c|}{} & \multicolumn{1}{c|}{start\_token\_id} & 50258 \\
\multicolumn{1}{c|}{\multirow{-5}{*}{Decoder}} & \multicolumn{1}{c|}{eos\_token\_id} & 50257 \\ \hline
\end{tabular}
}
\caption{\label{tab:statistic_model}\textcolor{black}{Partial hyperparameters of the TellWhisper.}} 
\end{table}

\section{More Details of Results}
\label{appendix:case}
\subsection{Hyperparameter Selection}
\begin{figure}[t]
\centering
\centerline{
\includegraphics[width=1\linewidth]{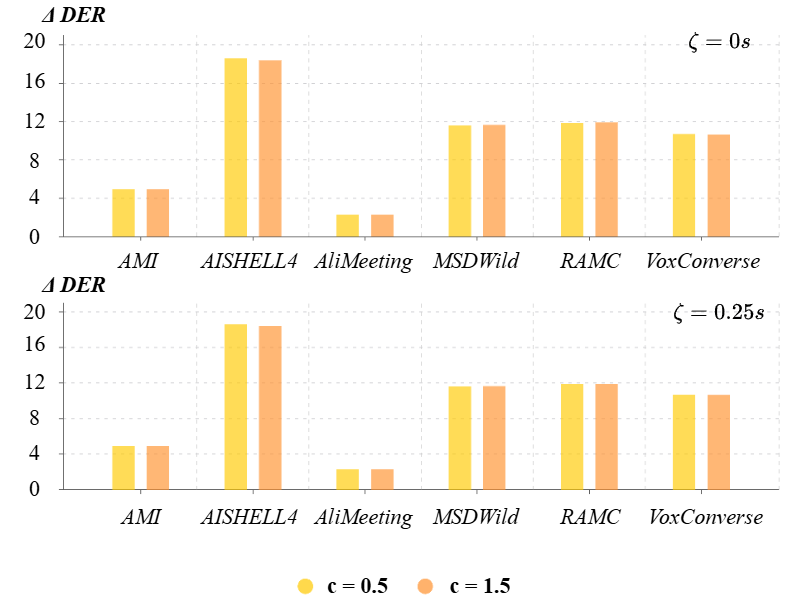}
}
\caption{Comparison of DER increases relative to $c=1.0$ under different hyperbolic-space negative-curvature parameter settings $c$ (collar($\zeta$) $=0$\,s / $0.25$\,s).}
\label{fig:hyper}
\end{figure}

Fig.\ref{fig:hyper} presents the change in DER induced by varying the hyperbolic curvature parameter $c$, measured against the default $c=1.0$ as $\Delta$DER $=\mathrm{DER}(c)-\mathrm{DER}(1.0)$, and compared under collar tolerances $\zeta\in\{0, 0.25\}$~s.
We observe that across six speaker diarization datasets, $c=1.0$ consistently yields the lowest DER under both collar settings. In contrast, $c=0.5$ and $c=1.5$ lead to uniform degradation on all datasets (i.e., $\Delta$DER is positive throughout).
In particular, the degradation is most pronounced on AISHELL4; MSDWild, VoxConverse, and RAMC also show large $\Delta$DER, suggesting that Hyper-SD is sensitive to curvature-related hyperparameters.
We attribute this trend to the joint influence of $c$ on the geometric properties of the hyperbolic manifold and its numerical behavior: under the commonly used Poincaré-ball parameterization, $c>0$ controls the magnitude of negative curvature and the distance scale (i.e., the degree of ``expansion'' of the space), and as $c \to 0$, the geometry gradually degenerates to Euclidean. 
Therefore, a smaller $c$ makes the space closer to Euclidean geometry, weakening the hyperbolic advantage in separating nearby classes (similar speaker representations), which may reduce inter-class/prototype separability; conversely, an excessively large $c$ increases curvature and makes distances more sensitive to position, especially near the ball boundary, thereby amplifying numerical errors and destabilizing manifold operations and optimization.
Overall, $c=1.0$ provides a better trade-off between representational capacity and optimization stability, and we therefore use $c=1.0$ as the default in Hyper-SD.

\balance

\subsection{Case Study}
\subsubsection{TellWhisper Performance on Overlapping Speech}
\begin{figure*}[h]
\centering
\centerline{
\includegraphics[width=1\linewidth]{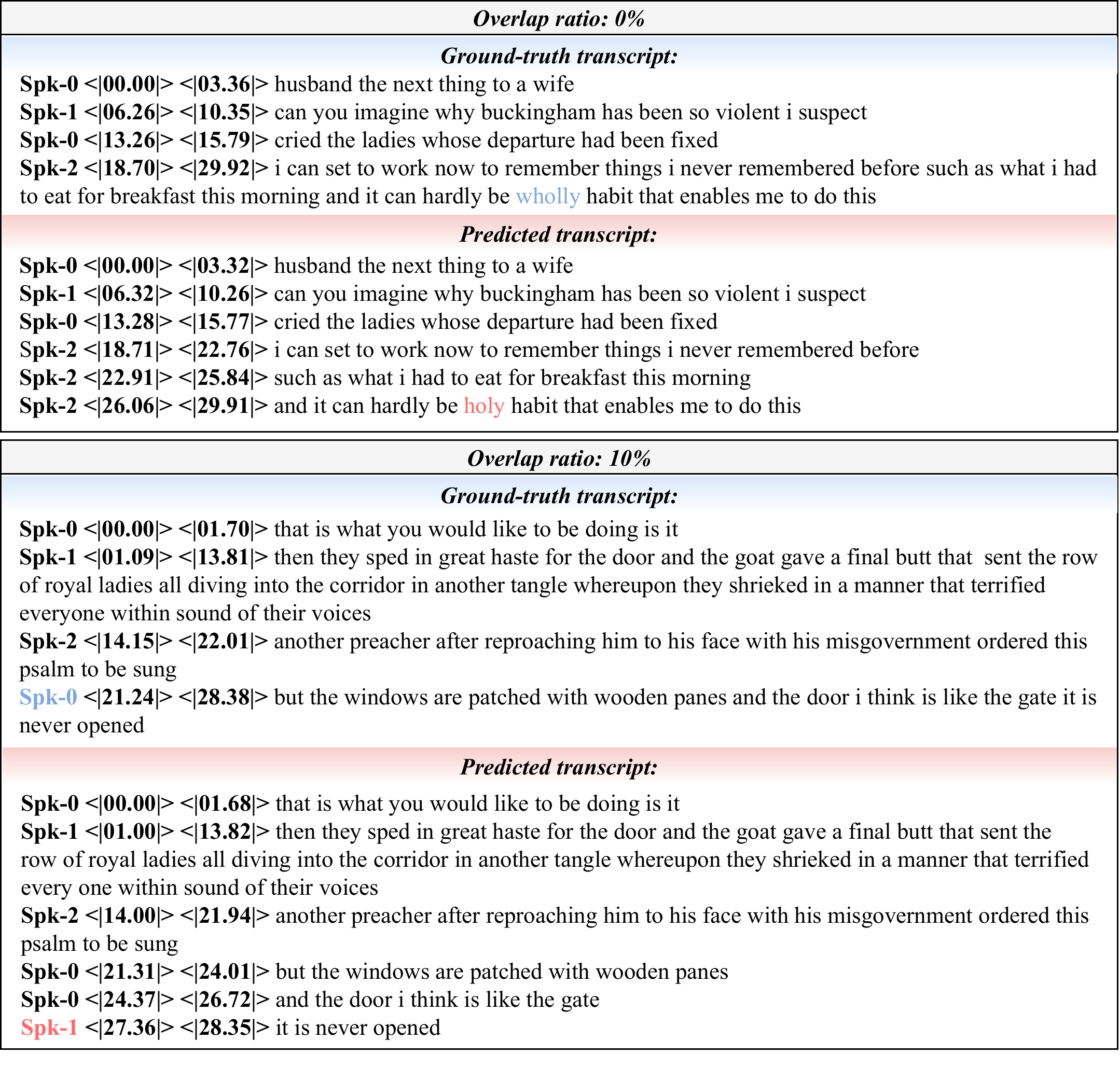}
}
\caption{Example transcripts on LibriCSS at 0\% and 10\% overlap.}
\label{fig:appdex-1}
\end{figure*}

\begin{figure*}[!t]
\centering
\centerline{
\includegraphics[width=1\linewidth]{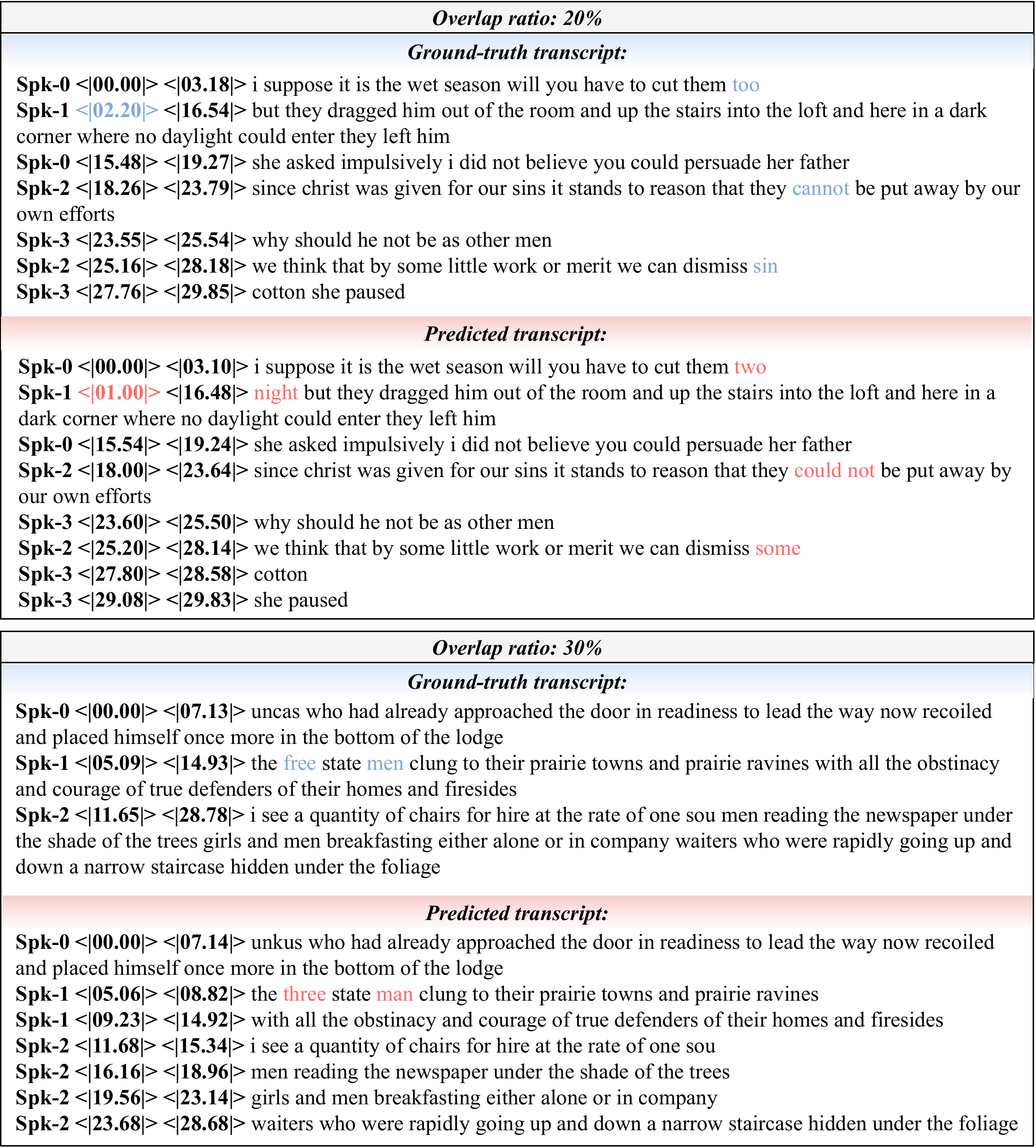}
}
\caption{Example transcripts on LibriCSS at 20\% and 30\% overlap.}
\label{fig:appdex-2}
\end{figure*}

As shown in Fig.~\ref{fig:appdex-1} and ~\ref{fig:appdex-2}, we conduct a qualitative case study on LibriCSS to examine model behavior under varying overlap ratios (0\%–30\%), focusing on speaker assignment, temporal alignment, and content transcription. 
Overall, TellWhisper remains robust as overlap increases and continues to produce coherent, well-structured outputs. 
In terms of content, the predicted transcripts largely preserve the semantics of the ground truth, with mismatches typically limited to occasional word-level substitutions in highly overlapped regions. Regarding temporal alignment, the model generally provides reasonable start/end boundaries.
Higher overlap may lead to slightly finer-grained segmentation or minor boundary shifts, yet the overall timing remains well aligned. For speaker attribution, predictions are consistently accurate under low-to-moderate overlap, while the few confusions observed at higher overlap are mostly localized around overlap windows and do not substantially disrupt the global conversational structure. Taken together, these visualizations suggest that although heavy overlap increases local ambiguity, our TellWhisper maintains strong performance across speaker, time, and content dimensions, demonstrating good robustness under challenging multi-speaker conditions.

\end{document}